\documentclass[twocolumn, 10pt]{article}
\usepackage[utf8]{inputenc}
\usepackage{todonotes}
\usepackage[switch]{lineno}

\usepackage{courier}
\usepackage{listings}
\usepackage{titlesec}
\titlespacing*{\section}
{0pt}{1ex}{1ex}

\title{Enumerating Hardware--Software Splits with Program Rewriting}
\author{Gus Smith, Zachary Tatlock, and Luis Ceze\\
University of Washington}
\date{}

\begin{document}

\maketitle

\section{Introduction}

The Instruction Set Architecture (ISA) has long been an essential abstraction in the computing world.
Describing complex hardware behavior using a manageable set of instructions has largely decoupled computer architects from compiler engineers, allowing the fields to make significant progress independently.
During the reign of Moore's Law, this decoupled advancement allowed software to benefit from regular improvements in hardware. %, with little need for innovation across the ISA boundary. 
% TODO Joesph says that this hadd always been going on, but it increased when moore's law ended.
% heterogeneity, etc etc.
% So maybe don't be so hard-line about this
But now, as Moore's Law fails to hold, researchers have been increasingly searching for innovations which cross the ISA boundary and challenge the assumption of a rigid software--hardware border.
%But now, as Moore's Law fails to hold, researchers have been exploring innovations which challenge the rigid software--hardware border set up by the ISA.
% Examples?
%For example, accelerators for machine learning effectively expand the ISA of a machine...
% Automated generation of ISAs?

% TODO Luis V: something about design splace
Especially interesting is the idea of \textit{software--hardware codesign:} the process of designing software and hardware in lockstep, rather than solidifying the ISA and designing in isolation.
A core problem in software--hardware codesign is in the sheer size of the design space.
Without a set ISA to constrain the software--hardware interface, the design space explodes to all possible combinations of hardware and software, with all possible interfaces between them.

This work presents a strategy for managing the massive hardware--software design space within the constrained domain of machine learning inference workloads and accelerators.
%In this ongoing work, we develop ways to manage this exponential hardware--software design space within the constrained domain of machine learning workloads and accelerators.
We first propose EngineIR, a new language for representing machine learning hardware and software in a single program. 
Then, using equality graphs---a data structure from the compilers literature---we suggest a method for efficiently enumerating the design space by performing rewrites over our representation.
%Using Equality Graphs---a data structure for efficiently enumerating an exponential space of designs---we seek to represent the design space efficiently.
%From here, the problem becomes twofold: first, building a representation for machine learning hardware--software ``programs'', and second, encoding a set of hardware--software program transformations which allow us to enumerate the search space.

\section{Overview of Solution}

% TODO Do we really need this figure?
%\begin{figure}
%    \centering
%    \includegraphics[trim=2 0 0 2, clip, width=\columnwidth]{vta-placeholder.png}
%    \todo[inline]{VTA image is a placeholder for space reasons}
%    \caption{some figure showing a common structure of ml accelerators. this could be based on vta's block diagram, or nvdla's. tpu paper probably has one as well.}
%    \label{fig:ml-accelerators}
%\end{figure}

% background on ml workloads and accelerators

%our goal is to enumerate the potential hardware--software splits within a machine learning workload.

% TODO mention that this is for inference only
% TODO the fourth thing confused joseph, remove entirely?
Within the scope of this work, we view hardware-accelerated machine learning inference workloads as being comprised of three distinct \textit{hardware} or \textit{software} components.
First, at their base, the workloads are built on calls to fixed size kernels, such as a $16\times16$ matrix multiplication;
a machine learning accelerator accelerates workloads by providing finely tuned \textit{hardware engines} implementing these kernels.
Second, \textit{software schedules} expand fixed-size kernel calls to take arbitrary-sized input by using loops or parallelism to call kernels multiple times. 
Finally, some concept of \textit{storage hardware} carries intermediate values between kernel invocations.

To explore hardware--software splits, we begin with ML workloads written in Relay, which is the intermediate representation used by the TVM compiler\cite{roesch2018relay}.
% It's good for us because of how we plan to do this, but we haven't yet described how we'll do this.
%Relay represents workloads in a hardware-agnostic way.
%Tensors are the base datatype in Relay.
%Relay's operators are machine learning kernels such as convolution and activation functions. %, which take tensors and produce tensors.
%Relay provides control flow constructs such as if-then-else and match expressions.
%A Relay program is thus a series of operator calls on tensors, strung together with control flow.
Relay represents a machine learning workload as a series of kernel calls, but does not make explicit the underlying hardware and software components described above.

%\paragraph{EngineIR.}
From Relay, we lower to EngineIR, a Relay-like language which fully reifies the hardware engines, hardware storage buffers, and software schedules underlying Relay programs. %: one construct for reifying computational engines, and another for reifying on-chip storage.
%\subparagraph{Engines.}
EngineIR engines represent underlying computational hardware.
%An engine \textit{declaration} can be seen as an interface to the underlying Hardware Description Language (HDL) implementation of the engine.
% TODO it could be interpreted as parameters being fixed
% variable parameters? parameter variables?
An engine is declared with a set of parameters (\texttt{H}, \texttt{W}, \texttt{C}, and \texttt{K}, in Figure \ref{fig:engine-declaration}) corresponding to the parameters of the underlying hardware design.
Each usage of a Relay operator will be converted to a call to an EngineIR engine instantiation with concrete parameters.
During this lowering process, a software schedule will also be created, implementing the kernel using the underlying hardware engine.
%\subparagraph{Storage.}
Similarly, each converted call will be given an explicit storage buffer for its output.
%In representing storage, there will need to be some notion of sharing, as we do not want to instantiate a different memory for each intermediate piece of data.
%We hope to utilize the functional, strongly-typed nature of Relay to build program analyses which allow us to generate sound and tight estimations of how memory can be shared, allowing us to instantiate only the memory we need.
%Finally, EngineIR will preserve the software control flow constructs present in Relay.
\begin{figure}
    \begin{lstlisting}
    engine conv2d<uint H, uint W, 
                  uint C, uint K>
                 (Tensor data, Tensor kernel);
    
    // %0 = relay.nn.conv2d(data, kernel);
    for i in ...
      for j in ...
        buffer<...>[..i..j..] = 
            conv2d<16, 16, 3, 3>
                  (data[..i..j..], kernel);
    \end{lstlisting}
    \caption{An engine declaration, plus a concrete instantiation of the engine wrapped in a software schedule of nested \texttt{for} loops.}
    \label{fig:engine-declaration}
\end{figure}
Figure \ref{fig:engine-declaration} shows an engine declaration for a convolution engine, including parameters for the input size (height, width, channels, and kernel size, respectively).
Additionally, it shows a Relay \texttt{nn.conv2d} call being reified into a software schedule over a concrete engine and concrete storage.
%For example, a call to a \texttt{conv2d} becomes a series of nested \texttt{for} loops, looping over the data.

%\paragraph{Rewrites and Equality Graphs.}
Once a workload is lowered from Relay to EngineIR, we enumerate the space of functionally equivalent hardware--software designs using \textit{equality graphs}\cite{nelson1981techniques} (e-graphs).
E-graphs are able to represent an exponential number of equivalent programs efficiently;
this key feature is what makes our hardware--software search space manageable.
To enumerate the search space, we first transform the EngineIR program into its equivalent e-graph.
We then perform \textit{rewrites} over the e-graph.
These rewrites encode transformations which alter the hardware--software split of a program, but preserve functional equivalence.
\begin{figure}
    \centering
    \includegraphics[width=.78\columnwidth]{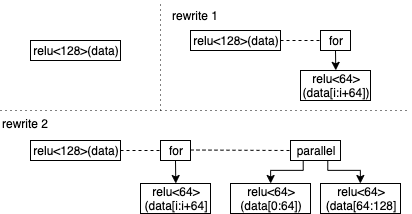}
    \caption{Example equality graph rewrites.}
    \label{fig:transform}
\end{figure}
Figure \ref{fig:transform} shows two such rewrites.
The program is a single call to a 128-bit wide ReLU, a common machine learning kernel.
Initially, the e-graph has a single node, representing a single usage of a 128-bit wide ReLU hardware engine.
Rewrite 1 encodes the knowledge that we can change the size of ReLU units in hardware by adding a software schedule which loops over the unit.
The e-graph is expanded with a new loop-based program; the dotted line indicates that the new program is equivalent to the old.
Rewrite 2 encodes the knowledge that we can parallelize a software for loop by instantiating more hardware.
% TODO add a line about "principled" exploration---only going towards profitable design, avoiding egraph blowup, etc
By incorporating a large body of such rewrites, and running them for a number of iterations, the e-graph will expand to include an exponential number of equivalent hardware--software programs.
When running an entire workload through a series of rewrites, we expect to see a wide range of design points represented.
For example, we should see designs which instantiate an engine for every kernel invocation, alongside designs which use complex software schedules and very little hardware.
From here, we can attempt to extract promising design candidates; however, the extraction procedure is out of the scope of this early work.

% TODO To get "breadth"-- talk about what a whole design looks like in the end. do we have a bunch of hardware engines? does each operator get its own engine?

% Describe the search problems
% Search problem can be implemented with AutoTVM-style ML tactics?
% Describe objective function/cost models
% Need cost models which are quick to evaluate

%% Evaluation methodology
\section{Evaluation Methodology}
The goal of this work is to provide a strategy for enumerating a massive design space.
As such, the evaluation will focus on our ability to generate a \textit{diverse} set of \textit{useful} designs.
A diverse set of designs should include many design points which differ significantly from each other.
However, this does not help in design space exploration if the designs themselves are not worth exploring.
Thus, the set of designs should also include many useful design points;
that is, designs which could turn into efficient hardware.

%% Related work
\section{Related Work}
%Equality graphs were first described in \cite{nelson1981techniques}, and have been used for a variety of design space enumerations, such as enumerating all possible assembly code implementations of a routine\cite{joshi2002denali}. 

There are many recent examples of automated machine learning hardware generation, such as \cite{hadjis2019tensorflow}, which demonstrates compilation from TensorFlow to FPGAs.
Their solution to the hardware--software split is to instantiate one engine for each type of kernel in the workload.
While this produces competent designs, the goal of this work is to allow for the easy enumeration and exploration of more complex (but potentially more profitable) splits.

\clearpage

\bibliographystyle{unsrt}
\bibliography{bib}

\begin{thebibliography}{1}

\bibitem{roesch2018relay}
Jared Roesch, Steven Lyubomirsky, Logan Weber, Josh Pollock, Marisa Kirisame,
  Tianqi Chen, and Zachary Tatlock.
\newblock Relay: a new ir for machine learning frameworks.
\newblock {\em Proceedings of the 2nd ACM SIGPLAN International Workshop on
  Machine Learning and Programming Languages - MAPL 2018}, 2018.

\bibitem{nelson1981techniques}
Charles~Gregory Nelson.
\newblock {\em Techniques for program verification}.

\bibitem{hadjis2019tensorflow}
Stefan Hadjis and Kunle Olukotun.
\newblock Tensorflow to cloud fpgas: Tradeoffs for accelerating deep neural
  networks.
\newblock In {\em 2019 29th International Conference on Field Programmable
  Logic and Applications (FPL)}, pages 360--366. IEEE, 2019.

\end{thebibliography}

\end{document}